\documentclass[12pt]{article}
\pdfoutput=1
\usepackage{amsmath}
\usepackage{graphics,graphicx,xcolor}

\def\be{\begin{equation}}
\def\ee{\end{equation}}
\def\arr{\begin{array}{rll}}
\def\ea{\end{array}}
\def\bea{\begin{eqnarray}}
\def\eea{\end{eqnarray}}

\def\N2{$N{=}2$}

\def\>{\rangle}
\def\<{\langle}
\def\+{\dagger}
\def\={\ =\ }

\def\bal{\begin{aligned}}
\def\eal{\end{aligned}}

\begin{document}
\begin{titlepage}
\setcounter{page}{0}
\begin{center}
{\LARGE\bf  A simple Lagrangian for shallow water }\\
\vskip 0.5cm
{\LARGE\bf  equations with constant potential vorticity}\\
\vskip 1.5cm
\textrm{\Large Anton Galajinsky \ }
\vskip 0.7cm
{\it
Tomsk Polytechnic University, 634050 Tomsk, Lenin Ave. 30, Russia} \\
\vskip 0.2cm
{\it
Tomsk State University of Control Systems and Radioelectronics,\\
Lenin ave. 40, 634050 Tomsk, Russia} \\

\vskip 0.2cm
{e-mail: galajin@tpu.ru}
\vskip 0.5cm
\end{center}

\begin{abstract} \noindent
Inspired by a recent gauge field description of shallow water equations 
[SciPost Phys. 14 (2023) 102], in this work we build a simple Lagrangian formulation for the
case of constant potential vorticity. An attractive feature of the construction is that
it does not appeal to a Clebsch-like parametrization or the use of auxiliary fields. It is also 
emphasized that
on the equator, where the Coriolis parameter vanishes, the shallow water equations exhibit extra
$SO(2,1)$ conformal invariance. Using the group-theoretic method, a new exact solution is found, which is 
associated with the scaling
symmetry transformation entering the conformal group.

\end{abstract}

\vspace{0.5cm}

PACS: 47.10.A, 47.10.Df, 47.54.Bd, 47.27.ef, 47.10.ab, 11.30.-j, 02.20.Sv \\ \indent
Keywords: shallow water equations, potential vorticity, Lagrangian formulation, conformal symmetry
\end{titlepage}
\renewcommand{\thefootnote}{\arabic{footnote}}
\setcounter{footnote}0

\noindent
{\bf 1. Introduction}\\

From a theoretical standpoint, an attractive feature of various hydrodynamic equations is that they admit
a simple Hamiltonian formulation. For example, the perfect fluid equations 
\be\label{PF}
\partial_t \rho + \partial_i ( \rho \upsilon_i )=0, \qquad
\rho \left(\partial_t+\upsilon_j \partial_j \right)\upsilon_i=-\partial_i p,
\ee
where $\rho(t,{\bf x})$ is fluid's density, $\upsilon_i (t,{\bf x})$ is velocity vector field, with $i=1,\dots,n$,
$p(t,{\bf x})$ is pressure obeying a specific equation of state $p=p(\rho)$, $\partial_t=\frac{\partial}{\partial t}$,
$\partial_i=\frac{\partial}{\partial x^i}$, and ${\bf x}$ designates a position vector in $n$-dimensional space, 
follow from the Hamiltonian
\be\label{H1}
H=\int d^n x \left(\frac 12 \rho \upsilon_i \upsilon_i+V(\rho) \right),
\ee
where the potential function $V(\rho)$ links to the pressure 
via the Legendre transform $p(\rho)=\rho V'(\rho)-V(\rho)$, provided the (noncanonical) Poisson brackets
\be\label{BR1}
\{\rho(t,{\bf x}),\upsilon_i (t,{\bf x}') \}=-\partial_i \delta^{(n)}({\bf x}-{\bf x}'), \quad 
\{\upsilon_i (t,{\bf x}),\upsilon_j (t,{\bf x}') \}=\frac{1}{\rho} 
\left(\partial_i \upsilon_j-\partial_j \upsilon_i  \right) \delta^{(n)}({\bf x}-{\bf x}')
\ee
are imposed, where $\delta^{(n)}({\bf x}-{\bf x}')$ is the Dirac delta function (for a review see \cite{JNPP}).

It appears quite natural to identify 
conserved energy with the Hamiltonian (\ref{H1}). A less trivial step is to find
(noncanonical) brackets obeying the
Jacobi identity (\ref{BR1}), that would reproduce the equations of motion (\ref{PF}).

Construction of a Lagrangian formulation for a given set of hydrodynamic equations may turn out to be much less
straightforward. In order to reproduce a desired Hamiltonian, one is usually 
forced to use either a Clebsch-like
parametrization of velocity vector field and deal with a constrained dynamics system\footnote{
For an elementary introduction to the subject of constrained dynamics see \cite{KS}.} 
in such a way that brackets
similar to (\ref{BR1}) show up as the Dirac brackets \cite{JNPP}, or introduce extra propagating fields (see e.g.
\cite{DT,TV} and the discussion below).

In certain physical situations, the use of a Lagrangian formulation may prove advantageous. 
This happens, 
for example, when particles constituting a fluid 
carry nonabelian charges or spin degrees of freedom \cite{BJLNP,NRR}, or if
the hydrodynamic limit of the AdS/CFT-correspondence is studied \cite{PSS},
or when the fluid-gravity correspondence is explored \cite{MR}.

Before proceeding further, we note a subtlety regarding the conventional 
Hamiltonian formulation of fluid mechanics.
When spatial dimension $n$ is even, the number of physical fields is odd. For example,
for $n=2$ the Hamiltonian formulation of the perfect fluid equations 
above suggests that the physical content is exhausted by
$\rho$ and $\upsilon_i$, with $i=1,2$, and hence momenta canonically conjugate to
$\rho$ and $\upsilon_i$ were previously removed from consideration as unphysical, 
yielding the Dirac bracket (\ref{BR1}). Thus, 
in two spatial dimensions 
the number of purported unphysical degrees of freedom is three. This, however, is at odds with the 
conventional constrained 
dynamics paradigm which states that unphysical phase space degrees of freedom always 
go away in even number \cite{KS}. This simple argument 
indicates that building a Lagrangian formulation for some hydrodynamic equations 
without either introducing extra 
propagating fields or imposing constraints reducing the number of physical degrees of freedom,
may turn out to be problematic in principle. 

In an interesting recent work \cite{DT}, a gauge field formulation has been proposed for 
shallow water 
equations\footnote{A brilliant account of the fascinating subject can be found in \cite{V}.}, 
including 
linearized approximation. 
An alternative Lagrangian for the linearized equations has been studied in \cite{TV}. 
The goal of this work is to 
reconsider the issue of constructing a Lagrangian formulation for shallow water equations. 
By applying the constrained dynamics analysis to the Lagrangians in \cite{DT,TV}, we demonstrate 
that both the formulations involve one more propagating field in addition to height and velocity vector field
characterizing shallow water. Then a simple Lagrangian formulation is proposed for shallow water equations
with constant potential vorticity. An attractive feature of the construction is that
it does not appeal to a Clebsch-like parametrization or the use of auxiliary fields. We also 
emphasize that
on the equator, where the Coriolis parameter is zero, the shallow water equations exhibit extra
$SO(2,1)$ conformal invariance. Using the group-theoretic method in \cite{PO}, a new exact 
solution is found, which is 
associated with the scaling
symmetry transformation entering the conformal group.

The organization of the work is as follows. In the next section, the structure of the shallow water equations is briefly reminded, including
the respective Hamiltonian formulation. The constrained dynamics analysis of a gauge field formulation in
\cite{DT} is implemented. It is shown that the model contains one more propagating field in addition
to height and velocity vector field characterizing shallow water. A similar 
constrained dynamics analysis of the linearized approximation in \cite{TV} is accomplished in sect. 3, 
leading to an analogous conclusion.
In sect. 4, a simple Lagrangian formulation for the shallow water equations with constant potential vorticity 
is built. Sect. 5 focuses on building exact solutions along the lines in \cite{PO}. 

Throughout the paper, summation over repeated indices is understood.

\vspace{0.5cm}

\noindent
{\bf 2. Constrained dynamics analysis of a gauge field formulation}\\

The shallow water equations 
\be\label{EE}
\partial_t h + \partial_i ( h \upsilon_i )=0, \qquad
\partial_t \upsilon_i-(f+\xi)\epsilon_{ij} \upsilon_j +\partial_i \left(\frac{\upsilon^2}{2}+g h\right)=0,
\ee
where $h(t,{\bf x})$ is the height of a fluid, $\upsilon_i (t,{\bf x})$ is the velocity 
vector field characterizing the horizontal flow,
$f$ is the Coriolis 
parameter\footnote{For the Earth, $f=2 \Omega \sin{\theta}$, where $\Omega$ is 
the Earth's angular 
rotation rate and $\theta$ is the latitude.}, $g$ is the gravitational acceleration, $\epsilon_{ij}$ is
the Levi-Civita antisymmetric symbol with $\epsilon_{12}=1$ and $\epsilon_{ij} \epsilon_{ik}=\delta_{jk}$, 
and $\xi=\epsilon_{ij} \partial_i \upsilon_j$ is fluid's 
vorticity\footnote{In view of the identity  
$\partial_i \upsilon_j-\partial_j \upsilon_i=\epsilon_{ij} \left(\epsilon_{kp} \partial_k \upsilon_p \right)$,
in two spatial dimensions
the vorticity tensor
reduces to the vorticity scalar $\xi=\epsilon_{ij} \partial_i\upsilon_j $.},
describe the dynamics of a fluid whose height is much smaller that its horizontal extent \cite{V}.

Leaving aside momentarily fascinating physical applications that include 
the dynamics of oceans and 
atmosphere \cite{V}, let us focus on the mathematical fact that (\ref{EE}) 
follow from the Hamiltonian
\be\label{HF1}
H=\frac 12 \int d^2 x \left( h \upsilon_i \upsilon_i+g h^2 \right),
\ee
provided the noncanonical Poisson brackets (which obey the Jacobi identities)
\be\label{BR}
\{h(t,{\bf x}),\upsilon_i (t,{\bf x}') \}=-\partial_i \delta^{(2)}({\bf x}-{\bf x}'),
\qquad
\{\upsilon_i (t,{\bf x}),\upsilon_j (t,{\bf x}') \}=\left(\frac{f+\xi}{h} \right) \epsilon_{ij} \delta^{(2)}({\bf x}-{\bf x}'),
\ee
are used in which $\delta^{(2)}({\bf x}-{\bf x}')$ is the Dirac delta function. 

The ratio $\frac{f+\xi}{h}$
entering the last bracket in (\ref{BR}) is known as the potential vorticity. 
Because it is materially conserved 
\be
(\partial_t+\upsilon_i \partial_i) \left(\frac{f+\xi}{h} \right)=0,
\ee
the potential vorticity provides an important characteristic of shallow water.
It will be of particular significance for the material in sect. 4.  

In a recent interesting work \cite{DT}, it was suggested that the action 
functional\footnote{In \cite{DT}, the action
is written in more compact and elegant form. The component action (\ref{CA}) is convenient 
for implementing the constrained dynamics analysis. }
\bea\label{CA}
&&
S=\int dt d^2 x \left(\frac{(\partial_t A_i-\partial_i A_0)(\partial_t A_i-\partial_i A_0)}{2(\partial_1 A_2-\partial_2 A_1)}
-\frac 12 g {(\partial_1 A_2-\partial_2 A_1)}^2+f A_0-A_0 \epsilon_{ij} \partial_i \beta \partial_j \alpha
\right.
\nonumber\\[2pt]
&&
\left.
\qquad \qquad \qquad  +
\partial_t \beta \epsilon_{ij} A_i \partial_j \alpha
-\partial_t \alpha \epsilon_{ij} A_i \partial_j \beta \right),
\eea
in which $A_{\mu}=(A_0,A_i)$, with $\mu=0,1,2$ and $i=1,2$, is an Abelian gauge field and $\alpha$ and $\beta$ are auxiliary fields,
provides a gauge theory formulation for the shallow water equations (\ref{EE}).
Let us implement the constrained dynamics analysis of (\ref{CA})
and demonstrate that the dynamical content of the model is actually larger.

Introducing momenta $p_0$, $p_i$, $p_\alpha$, $p_\beta$ canonically conjugate to the fields
$A_0$, $A_i$, $\alpha$, $\beta$, from the Lagrangian density in (\ref{CA}) one finds three primary constraints
\be\label{PC1}
p_\alpha=-\epsilon_{ij} A_i \partial_j \beta, \qquad p_\beta=\epsilon_{ij} A_i \partial_j \alpha,
\qquad p_0=0, 
\ee
as well as the equation which links the velocity $\partial_t A_i$ to the 
pairs underlying the canonical formalism
\be
\partial_t A_i=(\partial_1 A_2-\partial_2 A_1)p_i+
\partial_i A_0.
\ee
The corresponding canonical Hamiltonian reads
\bea\label{CH1}
&&
H=\int d^2 x \left( (p_\alpha+\epsilon_{ij} A_i \partial_j \beta) \lambda_\alpha+
(p_\beta-\epsilon_{ij} A_i \partial_j \alpha) \lambda_\beta+p_0 \lambda_0+\frac 12 
(\partial_1 A_2-\partial_2 A_1)p_i p_i
\right.
\nonumber\\[2pt]
&&
\qquad \qquad \qquad 
\left.
+p_i \partial_i A_0+\frac 12 g {(\partial_1 A_2-\partial_2 A_1)}^2-f A_0+A_0 \epsilon_{ij} \partial_i \beta
\partial_j \alpha \right),
\eea
where $\lambda_\alpha$, $\lambda_\beta$ and $\lambda_0$ are the Lagrange multipliers accompanying the
primary constraints (\ref{PC1}).
It is assumed that the canonical pairs $(\alpha,p_\alpha)$, $(\beta,p_\beta)$
and $(A_i,p_i)$ obey the conventional Poisson brackets
\bea
&&
\{\alpha(t,{\bf x}),p_\alpha (t,{\bf x}')\}=\delta^{(2)}({\bf x}-{\bf x}'), \qquad 
\{\beta(t,{\bf x}),p_\beta (t,{\bf x}')\}=\delta^{(2)}({\bf x}-{\bf x}'), 
\nonumber\\[2pt]
&&
\{A_i (t,{\bf x}),p_j (t,{\bf x}')\}=\delta_{ij} \delta^{(2)}({\bf x}-{\bf x}').
\nonumber
\eea

Conservation of the first two constraints in (\ref{PC1}) over time determines the 
Lagrange multipliers
$\lambda_\alpha$ and $\lambda_\beta$
\be
\lambda_\alpha=-\epsilon_{ij} p_i \partial_j \alpha, \qquad \lambda_\beta=-\epsilon_{ij} p_i \partial_j \beta,
\ee
meaning that $p_\alpha=-\epsilon_{ij} A_i \partial_j \beta$, 
$p_\beta=\epsilon_{ij} A_i \partial_j \alpha$ are second class constraints, while conservation of $p_0=0$
yields the secondary constraint
\be\label{SC1}
f+\partial_i p_i- \epsilon_{ij} \partial_i \beta \partial_j \alpha=0.
\ee

As the next step,
one verifies that the Poisson bracket of (\ref{SC1}) with the 
canonical Hamiltonian (\ref{CH1}) is zero and hence there are no tertiary constraints in the theory. 
The Dirac-Bergmann algorithm ends here. One can also check that (\ref{SC1}) Poisson commutes with
other constraints meaning that it is first class and likewise is $p_0=0$.

Now let us count physical degrees of freedom in the theory described by the action 
functional (\ref{CA}). According to the constrained dynamics nomenclature \cite{KS}, one second 
class constraint
removes one phase space degree of freedom from consideration, 
while one first class constraint eliminates
two phase space degrees of freedom. For the case at hand, the first class constraint 
$p_0=0$ renders the pair $(A_0,p_0)$ 
unphysical.\footnote{It is customary to impose the gauge fixing condition $A_0=0$.}
Likewise, two second class constraints $p_\alpha=-\epsilon_{ij} A_i \partial_j \beta$, 
$p_\beta=\epsilon_{ij} A_i \partial_j \alpha$, and one first class constraint $f+\partial_i p_i- \epsilon_{ij} \partial_i \beta \partial_j \alpha=0$
imply that the canonical pairs $(\alpha,p_\alpha)$, $(\beta,p_\beta)$ can be regarded as unphysical, leaving one
with four physical phase space variables $A_i$ and $p_i$ obeying the first order equations of motion
\be
\partial_t A_i=(\partial_1 A_2-\partial_2 A_1)p_i, \qquad
\partial_t p_i=(f+\partial_k p_k) \epsilon_{ij} p_j-
\epsilon_{ij} \partial_j \left(\frac 12 p_k p_k+g \epsilon_{kp} \partial_k A_p \right).
\ee

At this stage, it is straightforward to verify that introducing three new fields 
$h$ and $\upsilon_i$ in the following way
\be
h=\epsilon_{ij} \partial_i A_j, \qquad \upsilon_i=-\epsilon_{ij} p_j
\ee
one does reproduce the shallow water equations (\ref{EE}). However, the number of physical phase space 
degrees of freedom characterizing the gauge field model (\ref{CA}) 
was four meaning that,
apart from the shallow water dynamics, it describes one more propagating field, e.g. $\partial_i A_i$.

\vspace{0.5cm}

\newpage
\noindent
{\bf 3. The linearized approximation}\\

From a physical standpoint, the so called linearized approximation is of particular interest.
It arises when one expands about
an average height $H$
\be
h(t,{\bf x})=H+\eta (t,{\bf x})
\nonumber
\ee
and keeps in the original shallow water equations (\ref{EE}) only the terms linear in $\eta$ and $\upsilon_i$
\be\label{LA}
\partial_t \eta + H \partial_i \upsilon_i=0, \qquad
\partial_t \upsilon_i-f\epsilon_{ij} \upsilon_j +g \partial_i \eta=0.
\ee
The system admits many interesting 
solutions including the Kelvin and Poincar\'e waves \cite{V}. 

In a recent work \cite{TV}, the following action functional
\be\label{Ac2}
S=\frac 12 \int dt d^2 x \left( \partial_t A_i \partial_t A_i+f \epsilon_{ij} A_i \partial_t A_j -g H 
\xi^2\right),
\ee
where $\xi=\epsilon_{ij} \partial_i A_j$,
has been proposed as a candidate to describe (\ref{LA}). Let us demonstrate that, similarly to the gauge field 
formulation discussed
in the previous section, the dynamical content implied by (\ref{Ac2}) is larger than that described by 
the linearized shallow 
water equations because the model contains
one more propagating field in addition to $\eta$ and $\upsilon_i$.

In contrast to the linearized approximation of (\ref{CA}), the model (\ref{LA}) is free from constraints. Introducing
momenta $p_i$ canonically conjugate to $A_i$ in the usual way, one links the velocity $\partial_t A_i$ to the canonical pair 
$(A_i, p_i)$
\be\label{eq1}
\partial_t A_i=p_i+\frac{1}{2} f \epsilon_{ij} A_j.
\ee
The corresponding canonical Hamiltonian
\be
H=\int d^2 x \left(\frac 12 p_i p_i+\frac{1}{2} f \epsilon_{ij} p_i A_j+\frac{1}{8} f^2 A_i A_i+
\frac 12 g H  \xi^2\right),
\ee
gives rise to the equations of motion which include (\ref{eq1}) and
\be\label{eq2}
\partial_t p_i=\frac{1}{2} f \epsilon_{ij} p_j-
\frac{1}{4} f^2 A_i-g H \epsilon_{ij} \partial_j \xi,
\ee
where the conventional Poisson bracket
\be
\{A_i (t,{\bf x}),p_j (t,{\bf x}')\}=\delta_{ij} \delta^{(2)}({\bf x}-{\bf x}')
\nonumber
\ee
was used.
As the system is free from constraints, it describes four phase space degrees of freedom, each of which
is governed by a first order differential equation of motion displayed in (\ref{eq1}) and (\ref{eq2}).

Introducing three new fields \cite{TV}
\be
\eta=H \xi, \qquad \upsilon_i=-\epsilon_{ij} p_j+\frac{1}{2} f A_i,
\ee
where $H$ is identified with the average height of a fluid above, one does reproduce the linearized shallow water 
equations (\ref{LA}).
Yet, like in the gauge formulation discussed in the previous section, the number of physical degrees of freedom described by the action (\ref{Ac2})
is four and the system contains one more propagating field in addition to $\eta$ and $\upsilon_i$.

At this point, it is instructive to turn back to the Hamiltonian formulation (\ref{HF1}). 
It seems natural to treat the respective noncanonical Poisson brackets (\ref{BR}) as the Dirac brackets
originating from a Lagrangian formulation (of a dynamical system with constraints)
in the process of eliminating unphysical degrees of freedom from consideration. As is well known, 
second class constraints always go in pairs and
eliminate even number of unphysical phase space 
degrees of freedom. Each first class constraint is usually accompanied by a gauge fixing condition, similarly 
removing two unphysical phase space 
degrees of freedom. 

The bracket (\ref{BR}) suggests that the physical content of the model is exhaused by
$h$ and $\upsilon_i$, with $i=1,2$, and hence momenta canonically conjugate to
$h$ and $\upsilon_i$ were previously removed from consideration as unphysical. In two spatial dimensions 
the number of purported unphysical degrees of freedom is three. This, however, is at odds with the 
conventional constrained 
dynamics paradigm which states that unphysical phase space degrees of freedom always go away in even number \cite{KS}. 

This simple argument 
indicates that building a Lagrangian formulation for the shallow water equation without either introducing extra 
propagating fields or imposing constraints reducing the number of physical degrees of freedom,
may turn out to be problematic in principle. 

The Lagrangian formulations above explore the former possibility. 
In the next section, we study the situation in which
the number of physical degrees of freedom is reduced by imposing an extra condition linking 
height of a fluid to its vorticity.

\vspace{0.5cm}

\noindent
{\bf 4. A simple Lagrangian for the case of constant potential vorticity}\\

Having encountered difficulties in formulating a Lagrangian for
the shallow water equations without introducing
extra propagating fields, in this section we focus on a less general situation
in which the potential vorticity of the fluid is constant
\be
\frac{f+\xi}{h}=\mbox{const}.
\ee
As shown below, the case admits a very simple Lagrangian formulation.

In two spatial dimensions a Lagrangian density describing
a free first order dynamical system involving a single vector field $\upsilon_i (t,{\bf x})$ 
reads
\be
\frac 12 \epsilon_{ij} \upsilon_i \partial_t \upsilon_j,
\nonumber
\ee
where $\epsilon_{ij}$ is the Levi-Civita symbol.
Rewriting the second equation (\ref{EE}) in the equivalent form
\be\label{EE1}
\epsilon_{ij} \partial_t \upsilon_j+(f+\xi)\upsilon_i +\epsilon_{ij} \partial_j 
\left(\frac{\upsilon^2}{2}+g h\right)=0,
\ee
one can easily build a Lagrangian density
\be\label{LD1}
\frac 12 \epsilon_{ij} \upsilon_i \partial_t \upsilon_j+
\frac 12 (f+\xi) \upsilon^2 +g h \xi,
\ee
which reproduces (\ref{EE1}) when the corresponding action functional is 
varied with respect to $\upsilon_i$.

Yet, a variation of the 
action with respect to $h$ reduces the problem to irrotational fluids only, 
$\xi=0$. This drawback can be circumvented by adding to (\ref{LD1}) an arbitrary function 
of $h$ and fixing the latter so as to produce a reasonable equation for $h$. 
In this work, we choose to add
\be
f g h-\frac 12 \alpha g h^2,
\ee
where $\alpha$ is a constant of dimension $s^{-1} m^{-1}$,
which yields
\be\label{eomh}
\alpha h=f+\xi
\ee
when the resulting action functional\footnote{As usual, an 
action functional is defined up to a (dimensionful) constant.}
\be\label{act2}
S=\int dt d^2 x \left(\frac 12 \epsilon_{ij} \upsilon_i \partial_t \upsilon_j+
(f+\xi) \left(\frac 12 \upsilon^2+g h \right) -\frac 12 \alpha g h^2\right)
\ee
is varied with respect to $h$. 
From (\ref{eomh}) it follows that the potential vorticity
$\frac{f+\xi}{h}$ is constant for the case at hand. Note also that
$h$ determined by (\ref{eomh}) automatically solves
the continuity equation
\be
\partial_t h+\partial_i \left( h \upsilon_i \right)=0,
\ee
provided $\upsilon_i$ obeys the Euler equation
\be\label{EE2}
\partial_t \upsilon_i-(f+\xi)\epsilon_{ij} \upsilon_j +
\partial_i \left(\frac{\upsilon^2}{2}+\frac{g(f+\xi)}{\alpha}\right)=0.
\ee

To summarize, (\ref{act2}) provides a simple action functional for
the shallow water equations with constant potential vorticity. A pleasant feature of
the formulation is that it does not appeal to a Clebsch-like parametrization or the use of
auxiliary fields.

Let us implement the constrained dynamics analysis of (\ref{act2}).
Introducing momenta $p_h$, $p_i$ canonically conjugate to $h$ and $\upsilon_i$, one 
finds three primary constraints
\be
p_h=0, \qquad p_i+\frac 12 \epsilon_{ij} \upsilon_j=0
\ee
and the canonical Hamiltonian
\be
H=\int d^2 x \left(p_h \lambda_h+\left(p_i+\frac 12 \epsilon_{ij} \upsilon_j \right) \lambda_i -
(f+\xi) \left(\frac 12 \upsilon^2+g h \right) +\frac 12 \alpha g h^2
\right),
\ee
where $\lambda_h$ and $\lambda_i$ are the Lagrange multipliers accompanying the primary constraints.
It is assumed that the canonical pairs $(h,p_h)$ and $(\upsilon_i,p_i)$ obey the conventional Poisson brackets
\be
\{h(t,{\bf x}),p_h (t,{\bf x}')\}=\delta^{(2)}({\bf x}-{\bf x}'), \qquad 
\{\upsilon_i (t,{\bf x}),p_j (t,{\bf x}')\}=\delta_{ij} \delta^{(2)}({\bf x}-{\bf x}').
\ee

Conservation of $p_h=0$ over time gives the secondary constraint
\be
\alpha h-(f+\xi)=0,
\ee
in which one recognizes one of the Lagrangian equations of motion (\ref{eomh}), whereas 
conservation of
$p_i+\frac 12 \epsilon_{ij} \upsilon_j=0$ fixes the Lagrange multiplier $\lambda_i$
\be
\lambda_i=(f+\xi) \epsilon_{ij} \upsilon_j -\partial_i \left(\frac 12 \upsilon^2+g h \right),
\ee
meaning that the pair of constraints $p_i+\frac 12 \epsilon_{ij} \upsilon_j=0$, with $i=1,2$, are second class.

Conservation of the secondary constraint $\alpha h-(f+\xi)=0$ over time determines the Lagrange
multiplier $\lambda_h$
\be\label{lmh}
\lambda_h=\frac{1}{\alpha} \epsilon_{ij} \partial_i \lambda_j=-\partial_i 
\left(\frac{(f+\xi)}{\alpha} \upsilon_i \right)=-\partial_i \left(h \upsilon_i \right),
\ee
implying that the pair of constraints $p_h=0$ and $\alpha h-(f+\xi)=0$ are second class. Thus, no tertiary constraints
appear and the Dirac-Bergmann algorithm ends here. Note that (\ref{lmh}) also ensures that $h$ obeys
the continuity equation
\be
\partial_t h=\{h,H \}=\lambda_h=-\partial_i \left(h \upsilon_i \right).
\ee

Finally, introducing the conventional Dirac bracket associated with the second class constraints and
using them in order to eliminate $(h,p_h)$ and $p_i$ from consideration, 
one gets the reduced Hamiltonian\footnote{In order to assign to $H_{red}$ the dimension of energy, 
one could multiply (\ref{Hred}) with a dimensionful constant and include the 
latter in the inverse power to the right hand side of (\ref{DB1}).}
\be\label{Hred}
H_{red}=\int d^2 x (f+\xi) \left(\frac 12 \upsilon^2+\frac{g(f+\xi)}{2\alpha} \right)
\ee
and the Dirac bracket
\be\label{DB1}
{\{\upsilon_i (t,{\bf x}),\upsilon_j (t,{\bf x}')\}}_D=\epsilon_{ij} \delta^{(2)}({\bf x}-{\bf x}'),
\ee
which specify the dynamics of the
remaining independent field $\upsilon_i$ 
\be
\partial_t \upsilon_i={\{\upsilon_i, H_{red} \}}_D=
(f+\xi) \epsilon_{ij} \upsilon_j -
\partial_i \left(\frac{\upsilon^2}{2}+\frac{g(f+\xi)}{\alpha}\right).
\ee
The latter correctly reproduces the Euler equation in (\ref{EE}) for the case of constant 
potential vorticity.

\vspace{0.5cm}

\noindent
{\bf 5. Exact solutions}\\

The shallow water equations (\ref{EE}) admit a plethora of physically relevant solutions \cite{V}.
In this section, we focus on the group-theoretic method in \cite{PO} which
builds a particular solution to partial differential equations in terms of 
variables and fields invariant under action of a given symmetry (sub)group.

Our first example is a rotating solution which originates from the following symmetry transformation of (\ref{EE})
\be\label{ST1}
t'=t, \qquad x'_i=x_i+\alpha_i (t), \qquad
h(t,{\bf x})=h' (t',{\bf x}'), \qquad \upsilon_i (t,{\bf x})=\upsilon'_i (t',{\bf x}')-\dot{\alpha}_i (t),
\ee
where the parameter function $\alpha_i (t)$ obeys the differential equation 
$\ddot{\alpha}_i-f \epsilon_{ij} \dot{\alpha}_j=0$, $f$ being the Coriolis parameter, implying
\be\label{ST2}
\alpha_1 (t)=\lambda_1 \cos{(ft)}+\lambda_2 \sin{(ft)}, \qquad
\alpha_2 (t)=-\lambda_1 \sin{(ft)}+\lambda_2 \cos{(ft)},
\ee
where
$\lambda_1$ and $\lambda_2$ are constants. When verifying the invariance of eqs. (\ref{EE})
under the transformation (\ref{ST1}), the following identities
\be\label{Id}
\frac{\partial}{\partial t}=\left(\frac{\partial t'}{\partial t}\right) \frac{\partial}{\partial t'}+\left(\frac{\partial x'_i}{\partial t} \right) \frac{\partial}{\partial x'_i},
\qquad \frac{\partial}{\partial x_i}=\left( \frac{\partial t'}{\partial x_i} \right) \frac{\partial}{\partial t'}+\left(\frac{\partial x'_j}{\partial x_i} \right)\frac{\partial}{\partial x'_j},
\ee
have been used.

Following the group-theoretic approach in \cite{PO}, one first introduces the generator of the symmetry 
transformation (\ref{ST1}), (\ref{ST2}) 
(for definiteness we focus on the $\lambda_1$-transformation and set $\lambda_2$ to zero)
\be\label{Gen}
\mathcal{D}=\cos{(ft)} \frac{\partial}{\partial x_1}-\sin{(ft)} \frac{\partial}{\partial x_2}
-f \sin{(ft)} \frac{\partial}{\partial \upsilon_1}
-f \cos{(ft)} \frac{\partial}{\partial \upsilon_2}.
\ee
Then one builds variables and fields which hold invariant under the transformation 
by switching to the characteristic equations\footnote{For a detailed account of 
the method of characteristics see \cite{SM}.} 
\be\label{CE}
\frac{d x_1}{\cos{(ft)}}=-\frac{d x_2}{\sin{(ft)}}=-\frac{d \upsilon_1}{f \sin{(ft)}}=-\frac{d \upsilon_2}{f \cos{(ft)}}
\ee
associated with the linear partial differential 
equation $\mathcal{D} A(t,x,h,\upsilon)=0$ with $\mathcal{D}$ in (\ref{Gen}) and an unknown $A(t,x,h,\upsilon)$.

Three first integrals of (\ref{CE}) 
\be
x_1 \sin{(ft)} +x_2 \cos{(ft)} =C_1, \qquad \upsilon_1-f x_2=C_2, \qquad \upsilon_2+f x_1=C_3
\ee
allow one to build the new variable $y$ and two new fields $u_{1,2} (y)$
\be
x_1 \sin{(ft)} +x_2 \cos{(ft)} :=y, \qquad \upsilon_1-f x_2:=u_1 (y), \qquad \upsilon_2+f x_1:=u_2 (y),
\ee
which hold invariant under the $\lambda_1$-transformation given above. 
Because the height transforms as a scalar, it seems natural to take it in the form $h=h(y)$. 

Rewriting the first equation in (\ref{EE}) in terms of the new
variable and fields, one gets $u_i (y)=\frac{\gamma_i}{h(y)}$, 
where $\gamma_i$ is a constant vector,
whereas the second equation in (\ref{EE}) gives
$\frac{\partial u_i}{\partial y}=0$ meaning that $h$ and $u_i$ are constants. 
The resulting solution
\be\label{SS1}
h=h_0, \qquad \upsilon_i=f \epsilon_{ij} x_j+\beta_i,
\ee
with constant $h_0$ and $\beta_i$, describes a variant of a steady flow which is characterized by a 
constant vorticity $\xi=-2 f$. As a trivial corollary, the potential vorticity is constant 
as well and the solution fits to illustrate the dynamical system studied in
the previous section.

Focusing on the dynamics of a liquid particle governed by the differential equation
\be\label{lp}
\dot{x}_i=\upsilon_i (t, x(t)),
\ee
with $\upsilon_i$ in (\ref{SS1}), one readily obtains the general solution
\be
\begin{pmatrix}
x_1 (t) \\
x_2 (t) \\
\end{pmatrix}=
\begin{pmatrix}
\cos{(ft)} & \sin{(ft)}\\
-\sin{(ft)} &\cos{(ft)} \\
\end{pmatrix} 
\begin{pmatrix}
\gamma_1 \\
\gamma_2\\
\end{pmatrix}+\begin{pmatrix}
\beta_1 t \\
\beta_2 t\\
\end{pmatrix},
\ee
where $\gamma_{1,2}$ and $\beta_{1,2}$ are constants,
which is a combination of a uniform rotation with constant angular velocity 
determined by the Coriolis parameter $f$ and a uniform
rectilinear motion.

Our second solution steams from the observation that at the equator, where the Coriolis parameter is zero, 
the shallow water equations hold invariant under 
extra $SO(2,1)$ group realized as follows\footnote{As was demonstrated in \cite{RS,HH1}, the perfect fluid equations
(\ref{PF}) accommodate extra $SO(2,1)$ symmetry, provided the polytropic equation of state
$p(\rho)=\nu \rho^{1+\frac{2}{d}}$, where $d$ is the spatial dimension and $\nu$ is a constant, is chosen. 
Setting $n=2$ in (\ref{PF}) and formally identifying $\rho$ with $h$ and $\nu$ with $\frac{g}{2}$, 
one gets precisely the shallow 
water equations with $f=0$, meaning that the latter exhibit the same symmetry group.}
\begin{align}\label{sl2}
&
t'=\frac{\alpha t+\beta}{\gamma t+\delta}, &&
x'_i={\left(\frac{\partial t'}{\partial t} \right)}^{\frac 12} x_i,
\nonumber\\[2pt]
&
h(t,{\bf x})= \left(\frac{\partial t'}{\partial t} \right) h' (t',{\bf x}'),
&&
\upsilon_i (t,{\bf x})={\left(\frac{\partial t'}{\partial t} \right)}^{\frac 12} 
\upsilon'_i (t',{\bf x}')
+\frac{\partial}{\partial t} {\left(\frac{\partial t'}{\partial t} \right)}^{-\frac 12} x'_i,
\end{align}
where the parameters obey the restriction $\alpha \delta-\beta \gamma=1$. For definiteness, below
we focus on the scaling symmetry transformation
\be
t'=e^\lambda t, \quad x'_i=e^{\frac{\lambda}{2}} x_i, \qquad
h(t,{\bf x})=e^\lambda  h' (t',{\bf x}'), \qquad \upsilon_i (t,{\bf x})=e^{\frac{\lambda}{2}} \upsilon'_i (t',{\bf x}'),
\ee
where $\lambda$ is a constant parameter. 

Introducing the generator of the dilatation
\be\label{D2}
\mathcal{D}=t \frac{\partial}{\partial t}+\frac 12 x_i \frac{\partial}{\partial x_i}-h \frac{\partial}{\partial h}-\frac 12
\upsilon_i \frac{\partial}{\partial \upsilon_i},
\ee
and the system of characteristic equations 
\be\label{CS1}
\frac{dt}{t}=\frac{2 d x_1}{x_1}=\frac{2 d x_2}{x_2}=-\frac{d h}{h}=-\frac{2 d \upsilon_1}{\upsilon_1}=-\frac{2 d \upsilon_2}{\upsilon_2},
\ee
associated with the linear partial differential 
equation $\mathcal{D} A(t,x,h,\upsilon)=0$, with $\mathcal{D}$ in (\ref{D2}) and an unknown $A(t,x,h,\upsilon)$,
one can use the first integrals of (\ref{CS1}) to build 
the scaling-invariant variable $y_i$ and two scaling-invariant fields $w(y)$ and $u_i (y)$
\be
\frac{x_i}{\sqrt{t}}:=y_i, \qquad 
t \cdot h(t,{\bf x})  :=w(y), \qquad \sqrt{t}  \cdot \upsilon_i (t,{\bf x}) := u_i (y).
\ee
In what follows we assume $t>0$.

Rewriting the first equation in (\ref{EE}) in terms of the scaling-invariant variables 
and fields, one gets
the following partial differential equation
\be
\frac{\partial}{\partial y_i} \left(w(y) \left(u_i (y)-\frac 12 y_i \right) \right).
\ee
Instead of solving it in full generality, 
we choose a simpler road and use it to fix 
the velocity vector field
\be\label{ups}
u_i (y)=\frac 12 y_i \qquad \Rightarrow \qquad \upsilon_i=\frac{x_i}{2 t},
\ee
which, after substitution to the second equation in (\ref{EE}), determines $w(y)$
and hence the height
\be\label{hh}
w(y)=h_0+\frac{x_i x_i}{8 g t} \qquad \Rightarrow \qquad h(t,{\bf x})=\frac{h_0}{t}+\frac{x_i x_i}{8 g t^2},
\ee
where $h_0$ is a constant.

It is easy to analyze dynamics of a liquid particle, which is governed by (\ref{lp}) with
$\upsilon_i$ in (\ref{ups})
\be
x_i (t)=c_i \sqrt{t}, \qquad \upsilon_i (t,x(t))=\frac{c_i}{2 \sqrt{t}}, \qquad
h(t,x(t))=\left(h_0+\frac{c_i c_i}{8 g} \right) \frac{1}{t},
\ee
where $c_i$ is a constant vector, meaning that $\upsilon_i (t,{\bf x}) $ and $h(t,{\bf x})$ 
in (\ref{ups}), (\ref{hh}) describe a surge of fluid that gradually spreads in all directions, 
decreasing in both height and speed. To the best of our knowledge, the solution (\ref{ups}), (\ref{hh}) is new.
The fact that on the equator, where the Coriolis parameter vanishes, the shallow water equations exhibit an 
extra $SO(2,1)$ conformal invariance appears to have escaped attention thus far. 

\vspace{0.5cm}

\noindent
{\bf 6. Conclusion}\\

To summarize, in this work the issue of constructing a Lagrangian formulation for the shallow water equations
was reconsidered. Constrained dynamics analysis of a gauge field formulation in \cite{DT}
and the linearized approximation in \cite{TV}, which was implemented in sect. 2 and 3, suggests that for building such 
a Lagrangian one should either 
introduce extra propagating fields or, alternatively, impose extra constraints 
removing some degrees of freedom from 
consideration. Linked to the latter option, a simple Lagrangian formulation
for the shallow water equations with constant potential vorticity was proposed, in which case 
height of a fluid was related 
to its vorticity. An attractive feature of the construction was that
it did not appeal to a Clebsch-like parametrization or the use of auxiliary fields. It was also 
emphasized that
on the equator, where the Coriolis parameter was zero, the shallow water equations exhibited extra
$SO(2,1)$ conformal invariance. Using the group-theoretic method in \cite{PO}, a new exact 
solution was found, which was 
associated with the scaling
symmetry transformation entering the conformal group.

There is a considerable body of recent 
work on perfect fluid dynamics
with nonrelativistic conformal symmetries, including construction of Hamiltonian and Lagrangian formulations, 
supersymmetrization,
and building exact solutions \cite{AG}-\cite{TS5} (for important earlier developments see
\cite{HH1,HH2,HZ}). It would be interesting to understand whether the shallow water equations might 
prove relevant when studying the hydrodynamic limit of the AdS/CFT-correspondence \cite{PSS}
or the fluid-gravity correspondence \cite{MR}.

\vspace{0.5cm}

\noindent{\bf Acknowledgements}\\

\noindent
This work was supported by the Russian Science Foundation, grant No 23-11-00002-Ext.

\end{document}